# Banach fixed-point between SEM image and EBSD diffraction pattern from a cylindrically symmetric rotating crystal

(B. Da[1,2*], L. Cheng[3], X. Liu[1,3], K. Shigeto[4]) contribute equally, K. Tsukagoshi[5#], T. Nabatame[5], Z. J. Ding[3,6†], Y. Sun[7], J. Hu[8], J. W. Liu[9], D.M. Tang[5], H. Zhang[2], Z. S. Gao[10], H. X. Guo[11], H. Yoshikawa[1,2], and S. Tanuma[2]

[1]*Research and Services Division of Materials Data and Integrated Systems, National Institute for Materials Science, 1-1 Namiki, Tsukuba, Ibaraki 305-0044, Japan*

[2]*Research Center for Advanced Measurement and Characterization, National Institute for Materials Science, 1-1 Namiki, Tsukuba, Ibaraki 305-0044, Japan*

[3]*Department of Physics, University of Science and Technology of China, Hefei, Auhui 230026, China*

[4]*Hitachi High-Tech Corporation, Hitachinaka, Ibaraki 312-8504, Japan*

[5]*International Center for Materials Nanoarchitectonics (WPI-MANA), National Institute for Materials Science, Tsukuba, Ibaraki 305-0044, Japan*

[6]*University of Science and Technology of China, Hefei, Anhui 230026, P.R. China*

[7]*Department of Applied Physics and Applied Mathematics, Columbia University, New York, NY, 10027, USA*

[8]*Department of Physics and Institute for Nanoscience and Engineering, University of Arkansas, Fayetteville, Arkansas 72701, United States*

[9]*Research Center for Functional Materials, National Institute for Materials Science, 1-1 Namiki, Tsukuba, Ibaraki 305-0044, Japan*

[10]*Institute of Electrical Engineering, Chinese Academy of Sciences, Beijing, 100190, P.R. China.*

[11]*Key Laboratory of MEMS of Ministry of Education, School of Electronic Science and Engineering, Southeast University, Nanjing, 210096, People's Republic of China.*

*DA.Bo@nims.go.jp

#TSUKAGOSHI.Kazuhito@nims.go.jp

†zjding@ustc.edu.cn

**Abstract**

The Kikuchi bands arise from Bragg diffraction of incoherent electrons scattered within a crystalline specimen and can be observed in both the transmission and reflection modes of scanning electron microscopy (SEM). Converging, rocking, or grazing incidence beams must be used to generate divergent electron sources to obtain the Kikuchi pattern. This paper report the observation of Kikuchi pattern from SEM images of an exceptional rotating crystal with continuous rotation in the local crystal direction and satisfying cylindrical symmetry, named a cylindrically symmetric rotating crystal. SEM images of cylindrically symmetric rotating crystals reflect the interactions between electrons and the sample in both the real- and momentum-space. Furthermore, we identify an unexpected mathematical relationship between the electron backscattered diffraction (EBSD) Kikuchi pattern matrix map and the SEM image of the present sample which can be rationalized as a concrete example of the Banach fixed-point theorems in the field of EBSD technique.

# Introduction

Scanning electron microscopy (SEM) is one of the most commonly used modern scientific instruments, producing images of a sample surface by scanning it with a focused electron beam [1]. The complex interaction between electron beam and sample surface produces various signals at each image pixel detected and displayed on a display unit that scans over the sample in parallel with the beam scanning. When the focused beam hits a focal point on the specimen, the signal intensity is measured by a detector to integrate the dwell time and is expressed as the pixel's brightness in a digital image. This gray-level image is related to particular characteristics of the specimen and can be interpreted as various image contrasts caused by different mechanisms [2,3]. Therefore, it is crucial to understand the mechanism of contrast and its numerical meaning in SEM. Any new contrast mechanism discovered in SEM has the potential to expand its range of applications and usefulness in research and industrial production.

The most common mode of SEM operation is the raster scan [4] of a specimen's surface, in which energetic electrons are focused into a beam with a small spread, and the beam is incident on the sample in the same direction and scanned point-by-point. With this parallel incident electron beam, the digital image obtained is a real space image, where each pixel in the image corresponds to a location in the real space of the sample.

Throughout the historical development of SEM and its applications, various contrast mechanisms in raster scanning mode have been understood. Electrons interact with the sample surface to produce electron signals containing various information about the sample surface, in which the sample's surface morphology and composition are essential [5,6]. The most common types of contrast are topographic contrast [7-10] and composition/elemental contrast [11-14], which are applicable and available for almost all specimens and provide the basis for SEM image formation.

In addition, some special contrast mechanisms, such as electric field [15,16], magnetic [17,18], electron charging [19-21], and plasmon gain [22-24] contrast exist in certain types of materials and are closely related to specific material properties. All these contrasts can be well-explained by the particle nature of electrons.

SEM also produces diffraction contrast [25-31], which usually reflects the specimen's structural information, such as grain orientation, local texture, phase identification and distribution. Only electrons' wave nature can explain the diffraction contrast's formation mechanism. To observe a relatively complete diffraction pattern of a crystalline or polycrystalline sample using an SEM, one needs to change the operating mode of the SEM.

Two modes of SEM operation are used to obtain relatively complete diffraction patterns. The first is to collect backscattered electrons from different directions exiting the sample surface when electrons are incident on the sample at a large angle. This mode is the electron backscattered diffraction (EBSD) mode [32], in which the incident electron beam enters the sample at a glancing angle and is scattered by the atoms inside the sample. Many of these, called backscattered electrons, escape from the sample surface because of multiple large angle scattering [33-35]. In the process of leaving the sample, the backscattered electrons that meet the Bragg diffraction condition [36], $2d\sin\theta = \lambda$, are diffracted from a particular family of crystal planes of the sample, forming two conical surfaces with the central axis perpendicular to the family of crystal planes. The two conical surfaces intersect with the receiving screen to form a bright band called the Kikuchi band. The centerline of each Kikuchi band corresponds to the intersection of the expanded crystal plane corresponding to this Bragg diffraction with the receiving screen [37]. Because each Kikuchi band is associated with Bragg diffraction from one side of a set of lattice planes, it has been an effective method for studying a substance's crystal structure [38] since it was observed by Kikuchi in transmission electron microscopy (TEM) as early as 1928 [39].

The second mode to observe a complete diffraction contrast is to scan the single crystal sample surface by changing the incidence direction of the electron beam so that it forms a continuously changing angle about the crystal direction of the sample [40]. Vibrating the electron beam causes it to diffract with the lattice at a particular location on the sample surface in different directions to form a Kikuchi pattern. This is the electron channeling pattern (ECP) mode [41]. The EBSD and ECP modes both form a Kikuchi pattern by continuously changing the angle between the emission or incident electrons and the sample's crystal orientation and subsequently analyzing the Kikuchi patterns to obtain the crystal structure information [42,43]. Therefore, the pixel position change of the acquired Kikuchi patterns in both EBSD and ECP modes corresponds to the momentum change of the electrons in the plane normal to the sample surface.

It is evident that the SEM raster scan mode is used to acquire information in the real space of the sample, while the ECP and EBSD modes are used to acquire information about the sample in the momentum space. It is an intriguing prospect that a sample could be investigated in both the real- and momentum-space from a single SEM image, but we have achieved this using a so-called cylindrical symmetric rotating crystal. In this sample, the crystal direction at a local position continuously rotates and satisfies cylindrical symmetry with respect to the central position. With this sample, we have observed both the contrasts of the surface morphology and a complete Kikuchi pattern in one SEM image measured in the raster scan mode. The contrast produced by the overlap of the real- and the momentum-space in this unique SEM image combines the particle and wave nature of electrons interacting with the sample bulk and surface. Subsequently, the fundamental reason for observing the Kikuchi pattern in the SEM images is revealed by analyzing the cylindrical symmetric rotating crystal using the EBSD technique. Finally, we point out that one Kikuchi pattern in the EBSD Kikuchi pattern matrix map is unique, having the "same" Kikuchi pattern as the SEM image observed using the cylindrical symmetric rotating crystal. In this case, the SEM image is a Banach fixed point of the EBSD Kikuchi map.

## Results

**Kikuchi pattern imaging mechanism in SEM.** We now consider the case shown in Fig. 1a, where the electron beam is scanned over a rotating crystal for which the local crystal orientation is rotated. As the electron beam is scanned over the sample, its angle with respect to the local crystal orientation changes, and therefore the backscattered electron signals collected by an SEM detector also change. Because $\theta = \theta_B$ (where $\theta_B$ is the Bragg angle) at points A and B, the angle between the incident electron beam and the local crystal direction is $\theta < \theta_B$ between these positions and collects enhanced backscattered electron signals. However, $\theta > \theta_B$ before and after A and B, so the backscattered electron signals decrease relative to their values at A and B.

If a complete two-dimensional raster is scanned over the crystal, as typically done in SEM, then the other lattice planes also contribute to the contrast, and the resulting signal map, the ECP, shows contrast bands from all planes or normal to the surface. The width of each band is twice the appropriate Bragg angle of the set of lattice planes from which it comes, and the angle between the bands is the angle between the corresponding sets of lattice planes. Thus, this ECP has lattice symmetry in the region being examined, and if a rotating crystal with cylindrical symmetry is scanned, it is possible to observe a Kikuchi pattern.

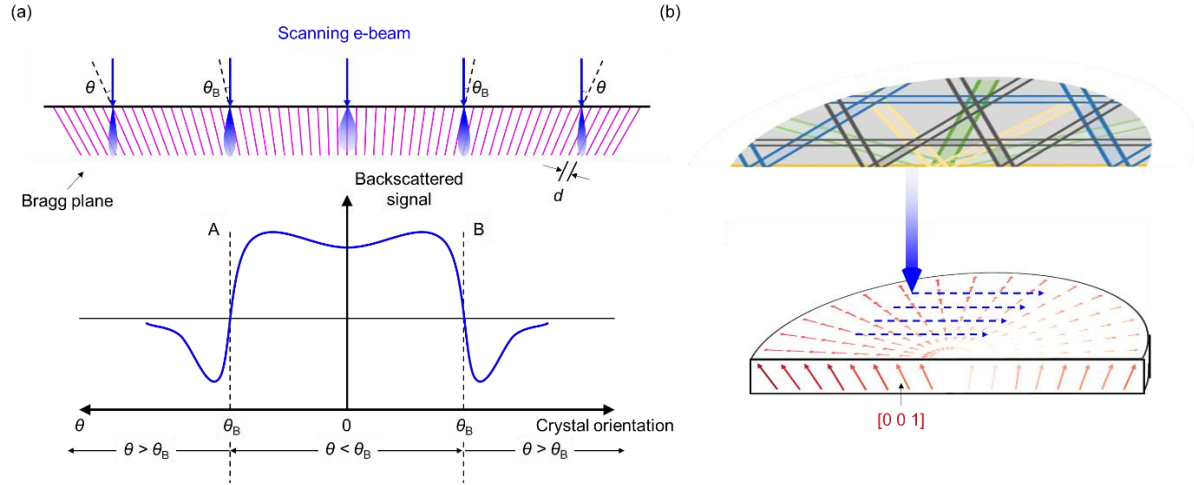

**Fig. 1 a** Geometric variation of the angle between the incident electron beam and the local crystal orientation while scanning a rotating crystal sample. During any line scan through the center of a rotating crystal, the angle $\theta$ between the incident electron beam and the local crystal orientation varies from greater than $\theta_B$ to less than $\theta_B$. The two symmetrical positions A and B are where $\theta = \theta_B$. This variation results in a change in the backscattered electron signal intensity during the line scan. **b** Schematic diagram of the crystal plane rotation of a rotating crystal and the corresponding Kikuchi pattern in the sample region.

Figure 1b shows a schematic of the cylindrically symmetric rotating crystal sample, in which the crystal's local orientation at different positions changes slowly and continuously. The relative local crystal orientation refers to the center of the sample satisfying cylindrical symmetry. As seen in Fig. 1b, the [001] crystal orientation deflection gradually increases from its normal position in the center of the crystal (perpendicular to the film plane) as the circular crystallographic front moves outward, as illustrated by the sketch of the [001] direction in the cross-section of the figure. In other words, during crystal growth, the unit cell rotates uniformly along the cross-section passing through the center point of the circular crystal film, and the unit cell (if traced along any radial outward direction of crystal growth) rotates permanently around an axis located in the plane of the film.

Therefore, the electron beam is scanned perpendicular to the sample's surface in the conventional SEM mode. The incident electrons diffract with the atomic lattice having different orientations at different landing positions, making it possible to observe the Kikuchi pattern in the entire sample scan image when the sample satisfies the special rotational structure described above. Although the contrast has been described in terms of the backscattered electron signal, any of the electron signals from the specimen (i.e., secondary, specimen current, or, if the sample is thin enough, the transmitted beam) show equivalent contrast features, and whichever is most convenient can be used.

**Observed SEM and EBSD images.** In this work, the rotating crystal is prepared based on bixbyite type $In_2O_3$. Figure 2a shows the crystallization of InSiO during the annealing process. The films grown on these substrates are partially crystallized, with a significant volume of material remaining amorphous. When the films are annealed at 300 °C, the amorphous films gradually crystallize, grow into approximately 1–2 μm diameter round crystal islands, and form distinct Kikuchi patterns. These rotating crystal islands are circular structures that grow quasi-isotropically and radially from individual nucleation sites.

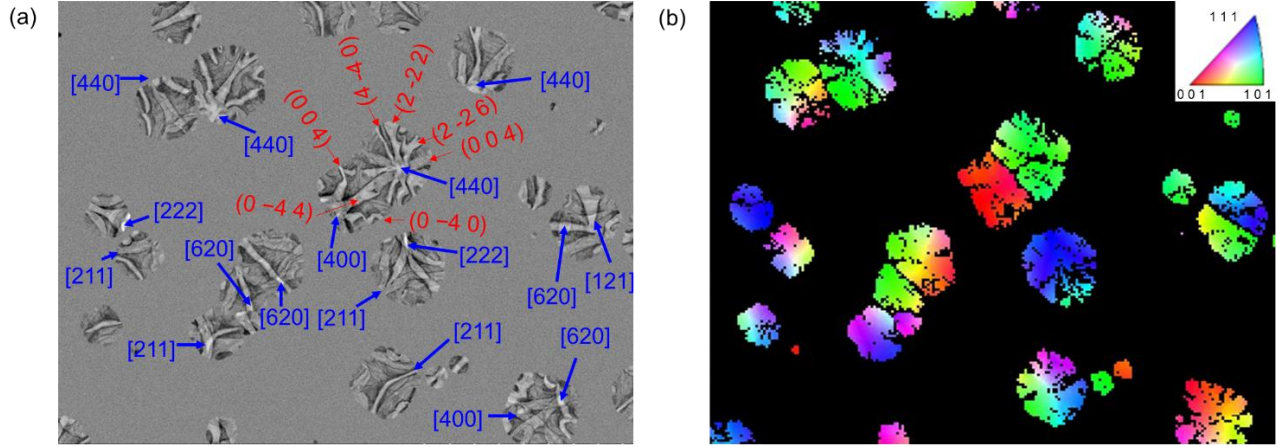

**Fig. 2 a** The observed SEM images of InSiO film detected by backscattered electrons during the crystallization process of amorphous film at 300 °C. **b** The observed EBSD-normal direction (ND) images in the same region.

It is reasonable to assume that these rotating crystal islands begin growing from randomly distributed nucleation sites in the amorphous film, followed by two-dimensional growth from the center. In some cases, the nucleation site is isolated. In other cases, the growth stops because of a neighboring crystalline site. This is a typical island nucleation and growth process in a thin film, as reported in Refs. [44-47]. These crystalline islands all exhibit distinct Kikuchi-like diffraction patterns in the SEM images. All the observed Kikuchi patterns belong to the cubic (bixbyite type) $In_2O_3$ structure. In addition, these crystalline islands exhibit different Kikuchi patterns in the SEM images, implying that each crystalline island's crystallographic orientation is different. The observed Kikuchi pattern has several characteristic patterns, which suggest several preferred nucleation directions in the crystallization process, similar to those observed in previous studies [48].

For example, crystallization usually starts with the formation of a crystalline nucleus with the $In_2O_3$ structure and with the [440] direction approximately normal to the film plane. The observed

Kikuchi patterns easily reveal the preferred lattice orientation with the [440] orientation. This orientation is deduced from the $In_2O_3$ crystal Kikuchi pattern, which allows indices to be assigned to the Kikuchi band (using the Kikuchi pattern spherical projection of $In_2O_3$). The nuclei always show strong lattice curvature, as indicated by the small inter-contour distances between the (hkl) and (-h-k-l) Kikuchi bands, always appearing in pairs. Six major Kikuchi bands are visible in the SEM image shown in Fig. 2a from the two sets of (2-26)-type planes: two sets of (2-22)-type planes, one (4-40)-type plane, and one (004)-type plane. Different sets of Kikuchi bands possess different bandwidths, intersecting in a "pole" and forming a bright, complex hexagon. Similarly, the Kikuchi patterns for the [400], [222] zone axis with different characteristics are shown in the SEM image. The indexes for these bands are also labeled in the SEM image.

These Kikuchi patterns can be assigned to the rotating crystal with zone axes of [211], [222], [400], [440], and [622], where the crystal planes are marked in Fig. 2a. We note that there is no Kikuchi pattern originating from the $SiO_2$ structure, indicating that the Si atoms are soluble in the $In_2O_3$ matrix at the rotating crystal region. This result is consistent with InSiO film crystallization information acquired by X-ray diffraction (XRD) [49], in which no XRD peak originated from crystalline $SiO_2$.

We note that the predicted inelastic mean free paths (IMFPs) of bixbyite $In_2O_3$ at an incident electron energy of 15 keV using the TPP-2M [50] and TPP-LASSO-S [51] empirical formulae are 16.1 and 16.5 nm, respectively. In addition, the Gaussian process regressor [68,69] machine learning model trained using well-established databases of solid material optical constants [52-58] and IMFP databases [59-67] resulted in a predicted bixbyite $In_2O_3$ IMFP of 15.9 nm at 15 keV incident electron energy. These predicted values are less than 1/3 of the InSiO film thickness, so it can be assumed that the contrast of the SEM images shown in Fig. 2a originates from the InSiO film.

The different colors in the EBSD-normal direction (ND) maps in Fig. 2b indicate the different orientations inside the rotating crystal islands. Orientation changes are visible within one island, and there is only one snowflake-like crystal "grain" without any significant crystal misorientation boundary in each rotational crystal island. The image analysis shows that the central area of every rotational crystal island spherulite has a homogeneous orientation rotational velocity, branching into larger features with a slightly different orientation rotational velocity upon further growth. However, it should be noted that such a rotational crystal is still a whole grain, which is entirely different from the previously reported ones [48,70] with spherulites that are spherical (or circular) structures consisting of fibers growing radially and quasi-isotopically from a single nucleation point. The orientation of the different island centers varies within a limited number of orientations, including [211], [222], [440], [400], and [622] zone axes, consistent with the observation from our XRD experiment [49].

**TEM cross-sectional images.** For further verification, the cross-sectional profile of the presented rotating crystal island for the [400] direction is provided in Fig. 3 using the TEM technique. The lattice structure of the sapphire substrate, as well as the crystalline region, is clearly seen to belong to the lattice structure of cubic bixbyite $In_2O_3$.

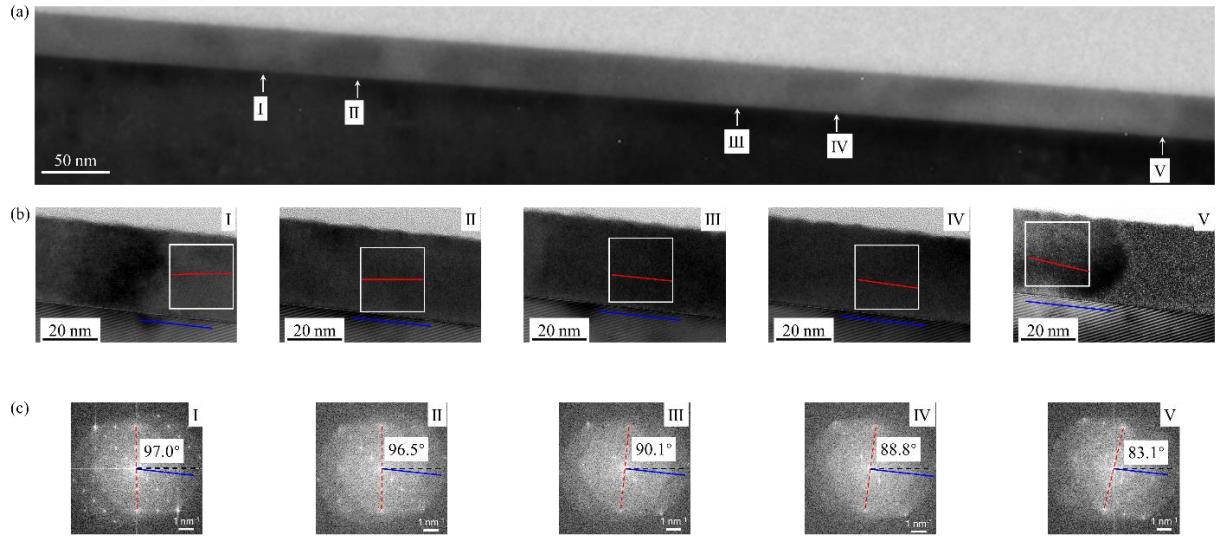

**Fig. 3 a** The cross-section of the rotating crystal island for the [111] direction, in which the central region is the rotating crystal island, and the neighboring two sides are the amorphous region. **b** Enlarged cross-sections of the crystal region in the TEM image. The blue lines are for the (100) plane in the sapphire substrate, and the red lines are for the (400) plane in InSiO. The lattice constants measured from the TEM images are 0.714 nm for InSiO and 0.435 nm for the sapphire substrate. **c** Fast Fourier transforms (FFTs) of the selected regions (marked by the dashed white rectangles in **b**) in the cross-sectional images. The blue lines are for the (100) plane in the sapphire substrate, the red dashed lines are for the [400] direction in momentum-space after FFT in InSiO, and the dashed black lines are for horizontal reference. The angles between the [400] direction in InSiO and (100) plane in the sapphire substrate are labeled in these images.

In these cross-sectional TEM images, the angle between the (400) plane in the InSiO cross-section (marked by the red line), and the (100) plane in sapphire (marked by the blue line), varies with the TEM image observation location. Fast Fourier transforms (FFTs) of these cross-sectional images are performed to demonstrate the variation of the (400) plane in the InSiO cross-section, as shown in the bottom panel (Fig. 3c). The red dashed lines are for the [400] direction in momentum-space after FFT, the blue lines are for the (100) plane in the sapphire substrate, and the black dashed lines are for horizontal reference.

The angles after FFT between the [400] direction of InSiO in momentum-space and the (100) plane in the sapphire substrate are measured and shown in Fig. 3c. The angle is 97.0 degrees when the observation location of the TEM image is at the leftmost edge of the rotating crystal island. As the observation location gradually moves to the right, the corresponding angle gradually decreases at an approximately constant rate, reaching 83.1 degrees when the observation location has moved to the rotating crystal island's rightmost edge (680 nm from the leftmost edge). The (400) crystal plane in the rotating crystal island clearly rotates continuously along the cross-sectional direction at a rotational speed of roughly 20.4 °/μm as the observation location changes in the rotating crystal island.

Additionally, the lattice constant of the rotating lattice island is determined by measuring the (400) lattice plane distance. The average lattice spacing of the (400) crystalline planes measured using these TEM cross-sectional images is 0.2513 nm. Thus, the lattice constant determined from this lattice spacing by the equation $a = d_{hkl}\sqrt{h^2 + k^2 + l^2}$ is 10.055 Å for rotating crystal islands. This value is almost identical to the value measured using the XRD technique (10.048 Å ± 0.018 Å).

However, the lattice constants of both measured rotating islands are slightly smaller than the original lattice constant of 10.094 Å ± 0.012 Å for polycrystalline $In_2O_3$ films prepared in the same way, indicating a 0.4% unit lattice shrinkage. Because tensile stresses usually expand the lattice parameters in the films, the shrinkage observed here is caused by Si dopants. Although a small amount of Si does not significantly change the crystal structure of cubic bixbyite, the lattice parameters of the rotating crystalline InSiO island differ significantly from those of pure $In_2O_3$.

The interface between the low-density amorphous InSiO and the high-density InSiO rotating crystal islands is observed in the TEM cross-sectional images of the two rotating island edge regions (Fig. 3). This sub-interface can also be seen as the crystallization front located at the film's middle depth. It is ahead of the crystallization front near the surface and the front near the substrate.

A significant amount of densification by shrinking must occur along the glass/crystal interface during crystallization when this front propagates as a whole parallel to the interface between the film sample and substrate. For thin films, the increase in density occurs preferentially in the direction perpendicular to the free surface. This directionality occurs because the change in film shape is unrestricted only in that direction [71] once the shrinkage rate along the highly diffuse crystalline front is slower than the propagation rate at the crystallographic interface. From the analogy of heterogeneous epitaxial crystal growth, it is expected that a type of dislocation is formed on the crystalline side of this interface, with mismatches appearing periodically to compensate for the lattice parameter mismatch [72]. An additional half-plane appears on the crystalline side because of the crystalline state's small interatomic distance (higher density).

According to speculation in Ref. [70], these unpaired dislocations initially appearing at the interface remain in the growing crystal volume. They may act as a built-in geometrically necessary dislocation (GND), which generates continuous lattice bending and is responsible for forming rotating crystal structures. A near-edge X-ray absorption fine structure (NEXAFS) measurement at the BL01B1 beamline of the SPring-8 synchrotron radiation facility was performed to verify this speculation [49]. The resulting NEXAFS spectra of the completely annealed rotating crystal InSiO sample demonstrate that the octahedral structure of the $InO_6$ units and tetrahedral $SiO_4$ units are preserved. The small number of tetrahedral $SiO_4$ units in the $InO_6$ octahedral network presumably causes local distortion in the bixbyite $In_2O_3$ matrix. These local distortions, in turn, become the GNDs necessary to rotate the InSiO thin-film crystal, which makes the InSiO film internally bend the crystal lattice planes around an axis lying in the film plane.

A similar phenomenon was found earlier for crystal growth in amorphous films of different substances [48,70-74], in which the local crystal orientation rotates uniformly along the cross-section that passes through the center point of the circular crystal film. Therefore, they are referred to as transrotational crystals. However, for the InSiO thin-film crystal used here, the relative local crystal orientation refers to the center of this sample possessing cylindrical symmetry in addition to continuous rotation in the local crystal direction. The details are discussed in the following section.

It is also notable that the observed interface propagation boundary differs slightly from the assumptions made in previous reports [70,71]. These previous reports assumed that the crystallization front near the surface leads to the crystallization front near the substrate. Here, the observed crystallization front near the surface and the crystallization front near the substrate both lag behind the crystallization front in the medium depth region of the thin InSiO film. This difference potentially leads to the rotating crystal, where the crystalline surface rotates not only along the radius direction, as shown in the TEM cross-sectional images, but also around the center of the circle. This results in a circular-symmetric pattern leading to the observed SEM Kikuchi pattern.

## Discussion

**Misorientation in rotating crystal island.** Figure 4a shows SEM and EBSD measurements of rotating crystal islands. The backscattered electron (BSE) images of the rotating crystal island show clear Kikuchi patterns, indicating that the central region of the rotating crystal island should be near the [440] zone axis. It is clear that six major visible Kikuchi bands arise from the two sets of (2-26)-type planes: two sets of (2-22)-type planes, one (4-40)-type plane, and one (00-4)-type plane. Different sets of Kikuchi bands possess different bandwidths, intersecting in a "pole" that forms a complex polygon composed of a bright central hexagon as the zone axis. This pole is the intersecting region of the (004), (2-26), and (-22-6) Kikuchi bands, as well as six adjacent triangles formed by the intersections of two of these three bands.

The sets of blurry lines (Fig. 4a) originating from the higher order reflection of the (-440) plane and flanking the major Kikuchi bands can be observed in the rotating crystal island with [440] direction as a rounded bright region with a clear dark outline. The intersections between these blurry lines and the major Kikuchi bands form some brighter nodes in the major Kikuchi bands. This is further confirmed by EBSD measurement. However, the EBSD map in the normal direction (EBSD-ND) is not homogeneous. Slightly varying colors in this map indicate the various slight local region misorientations of the rotating crystal island referring to the [440] crystal direction. In this case, the relative misorientation of the angle distribution (EBSD-angle), and the relative misorientation of the axis distribution (EBSD-axis), are plotted in Fig. 4a.

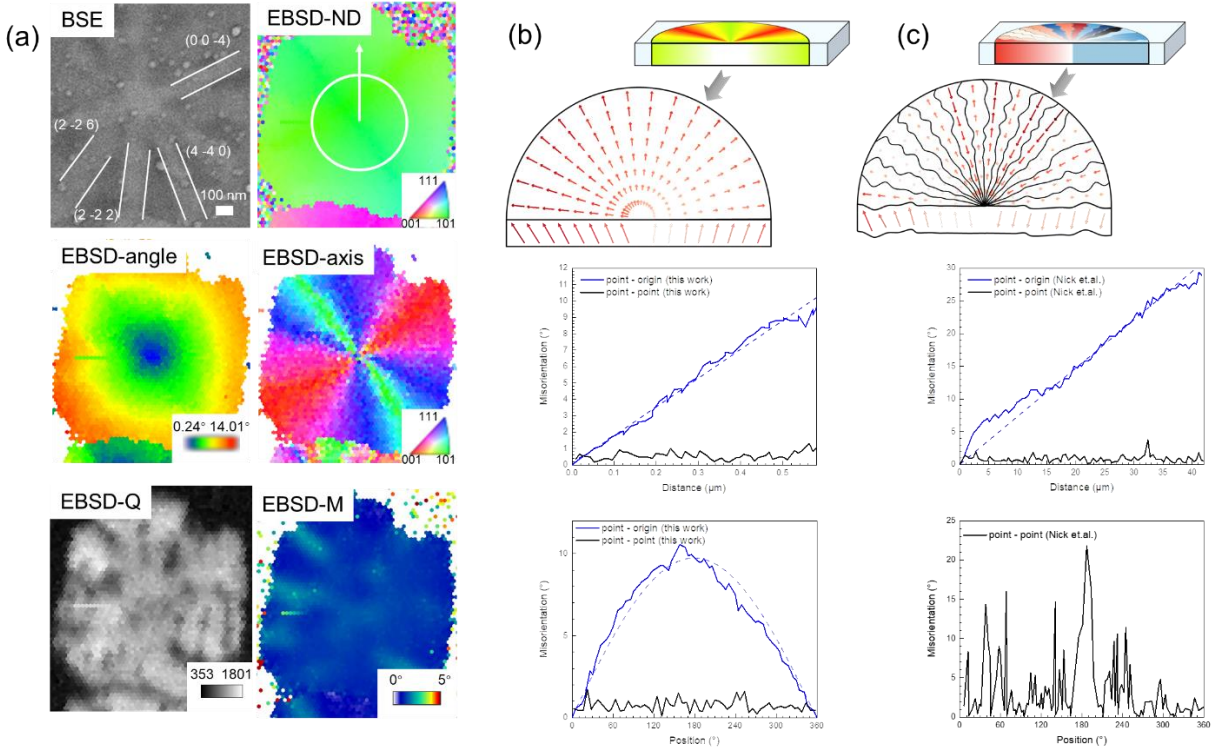

**Fig. 4 a** SEM and EBSD measurements of a rotating crystal island with a central region near the [440] zone axis. From top left to bottom right are the BSE image, EBSD map in the normal direction (EBSD-ND), EBSD relative misorientation-angle map (EBSD-angle), EBSD relative misorientation-axis map (EBSD-axis), EBSD image quality map (EBSD-Q), and EBSD misorientation map (EBSD-M), respectively. **b** Top: The schematic diagram of crystal direction rotation in an ideal rotating crystal island, where the direction of the red arrow represents the direction of any crystal plane. Middle: Crystal misorientation angle profiles measured along the white circle (panel **a**) around the center of the rotational crystallization island. The reference line for the theoretical calculation of point-origin misorientation around an ideal circular symmetry rotating crystal is shown as the blue dashed curve, determined from

$$\alpha = \arccos \frac{\vec{r}_{\text{origin}} \cdot \vec{r}_{\text{point}}}{\left|\vec{r}_{\text{origin}}\right| \cdot \left|\vec{r}_{\text{point}}\right|} = \arccos\left(\sin^2 \alpha_0 \cos\theta + \cos^2 \alpha_0\right)$$

, where $\alpha_0 = 4.86°$ as measured from the EBSD measurement. The inset is the sketch for calculating the corresponding point-origin misorientation from an ideal circular symmetry rotating crystal. Bottom: Crystal misorientation angle profiles measured along the white arrow from the center to the edge of the rotational crystallization island. **c** Top: The schematic diagram of crystal direction rotation in a previously observed rotating crystal island [70]. Middle: Crystal misorientation angle profiles measured along a circle around the center of the rotational crystallization island from Fig. 2 in Ref. [70]. Bottom: Crystal misorientation angle profiles measured along a radius from the center to the edge of the rotational crystallization island.

In the EBSD-angle map, different colors represent the relative out-of-plane direction between the local orientation and the average crystal direction of the entire crystallization region. The EBSD-angle map shows that the angle of the central region of the rotating crystal island nearly overlaps with the average crystal direction, which implies two-dimensional spherulitic crystal growth. Furthermore, the relative misorientation angle changes from 0° to 14° from the center to the edge, forming a roughly circular-symmetric distribution. This circular symmetry implies that the relative misorientation angle increases at a constant rate with the distance from the central region.

Different colors in the EBSD-axis map represent the relative in-plane direction differences between the local orientation and the averaged crystal direction of the island. The presented EBSD-axis map shows a significant 180º rotational symmetry pattern, meaning that the local crystal direction along any circle around the central region gradually and uniformly rotates by 360°. Furthermore, any diameter in the rotating crystal island has approximately the same color distribution. This signifies that the in-plane relative misorientation between the local orientation and average orientation of the entire crystallization region is approximately the same along each diameter.

Based on the EBSD-angle and EBSD-axis maps, the crystal plane in the rotating crystal island not only rotates uniformly along any radial outward crystal growth direction but also rotates permanently around an axis located at the rotating crystal island center. This is even clearer in the EBSD misorientation map shown in Fig. 4a (EBSD-M). The misorientation is almost the same on this rotating crystal island at about 2°, meaning that the crystal orientations of any pixel on this rotating crystal island and the surrounding neighboring pixels are almost equal. In other words, this disk-shaped rotating crystal island, with crystal orientations at any local position and along any adjacent direction, continuously rotates at a nearly uniform velocity.

EBSD image quality (EBSD-Q) plots are also presented in Fig. 4a. This snowflake-shaped crystal quality distribution indicates that the rotation of the crystal orientation originates from the density difference between the amorphous and crystalline InSiO films during crystallization. According to the EBSD-Q map, the central region of the rotating crystal island has a relatively uniform crystal quality, branching into larger features of decreasing quality after further growth and forming finger-like patterns at its front end.

At a certain distance from the core, more anisotropic growths appear with a clear fiber structure having finger patterns at their growth fronts. Two different mechanisms can usually cause such growth. The first mechanism is crystallographic branching [75], when one primary fiber grows from the nucleation center without forming small angle grain boundaries (< 3º). The second mechanism is the so-called non-crystallographic branching [76-79], when new sub-crystals (secondary fibers) with small misorientations (> 3º) heterogeneously nucleate on the side of an already growing fiber to grow as new fibers, primarily in radial directions.

The possibility of growing a rotating crystal arises when the two mechanisms described above are present during crystal formation. If the crystallographic branching effect dominates, it grows into radial spherical crystals with finger patterns, resulting in a snowflake-like crystal quality distribution, as observed in the EBSD-Q map. Conversely, if the non-crystallographic branching effect dominates, it grows into radial spherical crystals with many slender fibers, as reported in previous observations in Se [80-82], $Fe_2O_3$ [81-83], and $V_2O_3$ [84] fibers in spherulites, as well as $Cr_2O_3$ [84], $V_2O_3$ [74], $Ta_2O_5$ [85], Ge-Te, Tl-Se, Cd-Te alloys [86], and single-crystal Cu-Te alloys [87], among others [88].

The tops of Fig. 4b and 4c show the local crystal rotation distribution for an ideal rotating crystal in this paper and the previously reported rotating crystals, respectively. The present rotating crystalline film forms a special crystallization region in the shape of a disk with varying local crystal orientations because the crystallographic branching effect dominates the crystallization process. However, in this disk-shaped crystallization region, the local crystal orientation relative to the crystal center's orientation not only rotates in the radial crystal growth direction but also permanently rotates around the axis located at the crystal center with the same rotational velocity.

By contrast, the non-crystallographic branching effect dominates the crystallization growth process in the previously reported rotating crystal films, forming a crystallization region consisting of fibers. Its local crystal orientation changes only gradually with the growth of the fibers, rotating continuously in the radial growth direction. Because of the difference in crystal orientation between fibers, the local crystal orientation does not undergo any coherent rotational behavior in tangential direction.

The middle images in Fig. 4b and 4c show the lattice misorientation of the rotating crystals in this paper and those previously reported in Ref. [70], respectively, running at a constant radius along the white arrow from the center to the edge. The crystal misorientation curves for these two rotating crystal types show similar apparent linear behavior. From Fig. 4b and 4c, a linear fitting can obtain the gradient of the crystal rotation angle (the crystal rotation velocity along the selected radius). The crystal direction rotational velocity is 17.6 degree/μm for the present rotating crystal and 0.74 degree/μm for the rotating crystal in Ref. [70]. This large difference arises from the significantly different crystal sizes. The diameter of the present rotating crystal island is about 1.6 μm, while the diameter of the rotating crystal island in Ref. [70] is about 70 μm.

The bottom images in Fig. 4b and 4c show the crystal lattice misorientation profile along the white circle (Fig. 4a, EBSD-ND) around the crystal center for the rotating crystals in this paper and previously reported in Ref. [70], respectively. The point-to-point misorientation curves for the present and previously reported rotating crystals are entirely different. For the present rotating crystal, the point-to-point misorientation curve is a slightly fluctuating horizontal line close to 1°, and its maximum fluctuation is 2°. Note that the overall accumulated misorientation for a continuous rotation vector around the axis is 360° for one circle. Therefore, the average value of 1° of the point-to-point misorientation curve in the 0–360° range indicates that the misorientation observed along the white circle is solely contributed by local crystal direction rotation at an approximately constant velocity. In order words, the whole present rotating crystal island is composed of one primary fiber growth from the nucleation center in which the crystal orientation rotates without forming small angle grain boundaries.

For the previously reported rotating crystal, the point-to-point misorientation curve constantly and dramatically fluctuates, with an average value of 3.42° and a maximum misorientation of 22°. This result means the previously reported rotating crystal consisted of fibers growing radially from a single nucleation point. Therefore, there is no continuous crystal direction rotation along the white circle, only local crystal direction rotation occurring at varying velocities inside each fiber. The dramatic rise and fall observed in the point-to-point misorientation curve are contributed by the intense and random misorientation variations between neighboring fibers.

The point-origin misorientation curve of the present rotating crystal is shown in Fig. 4b to show how the local crystal orientation rotates along the white circle. The point-origin misorientation formed by a vector rotating continuously around the axis is also plotted as a reference curve in Fig. 4b and can be obtained from the following equations:

$$\vec{r}_{origin} = (r, 0, \frac{r}{\tan \alpha_0}); \vec{r}_{point} = (r \cos \theta, r \sin \theta, \frac{r}{\tan \alpha_0})$$

$$\alpha = \arccos \frac{\vec{r}_{origin} \cdot \vec{r}_{point}}{|\vec{r}_{origin}| \cdot |\vec{r}_{point}|} = \arccos \left( \sin^2 \alpha_0 \cos \theta + \cos^2 \alpha_0 \right)$$

where $\alpha_0$ is the mean misorientation relative to the rotation axis at radius $r$, and $\theta$ is the azimuth on the plane perpendicular to the rotation axis. The angle $\alpha_0$ between the vector and the rotation axis is 4.86° as measured from the EBSD-angle map. This reference point-origin misorientation curve may be regarded as the point-origin misorientation curve measured from an ideal circular symmetry rotating crystal, as shown in Fig.4a.

The point-origin misorientation curve for the present rotating crystal is consistent with that determined from a vector rotating continuously around the axis. In other words, the point-origin misorientation curve for the present rotating crystal is consistent with an ideal circular symmetry rotating crystal. In fact, the local crystal orientation in the present crystal rotates the same way along a circular trajectory for any radius once the center of the circular trajectory overlaps with the crystal's center. Therefore, using the rotating crystal center direction as a reference, the relative local crystal direction satisfies circular symmetry throughout the present rotating crystal region.

This additional circular symmetry is the essential difference between the present and previously reported rotating crystals and is also the fundamental reason for the observed Kikuchi pattern in the SEM images presented here.

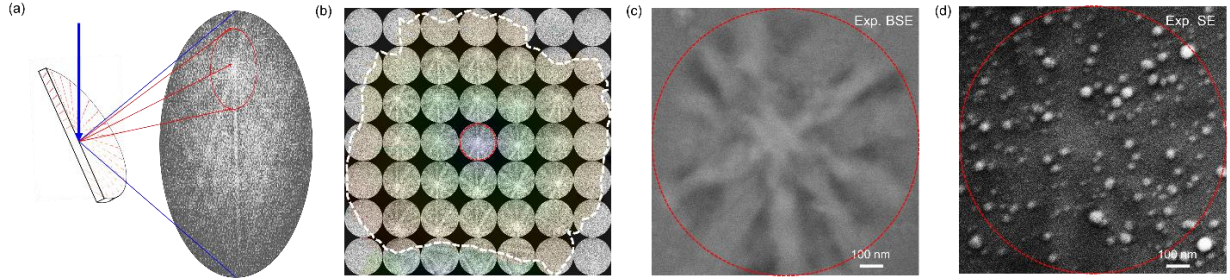

**Fig. 5 a** The geometry of the EBSD measurement. The central part of the Kikuchi pattern distribution matrix region was projected on a planar screen to a perpendicular surface. The Kikuchi pattern of the emitted electrons, within ±15° of the sample's surface normal, is coiled in the planar screen. **b** EBSD Kikuchi pattern matrix map observed by a charge-coupled device (CCD) camera in the EBSD measurement at intervals of 200 nm from the rotating crystal island sample, formed by electrons escaping within ±15° along the surface normal direction. Enlarged Kikuchi pattern region in the CCD image at the central region of the rotating crystal island. A dashed curve marks the rotating crystal island. The colors inside the island are following EBSD-angle map (Fig. 4a). A red circle marks the Kikuchi pattern corresponding to the Banach fixed point. **c** The measured BSE image of InSiO rotational crystal islands. **d** The measured secondary electron (SE) image of InSiO rotational crystal islands. The image's contrast is adjusted to highlight the Kikuchi pattern's fine structures.

**Banach fixed point in EBSD Kikuchi patterns.** As shown in Fig. 5a, the specimen is significantly tilted (about 70°) from the horizontal plane in the EBSD measurement geometry. Therefore, electrons escaping from the specimen along the surface normal form a Kikuchi pattern.

This pattern appears on the upper side of the center of the charge-coupled device (CCD) image at about 50°. The center of the CCD camera is on the same level as the center of the sample.

Figure 5b shows the EBSD Kikuchi pattern matrix maps observed in a selected region of the CCD camera at different electron landing positions in the rotating crystal sample. This result corresponds to the emitted electrons within ±15° of the sample's surface normal. The Kikuchi pattern is clear for electron landing positions within the boundary of the rotating crystal island (marked by the white dashed curve), and no pattern is found for electron landing positions in the amorphous region outside the rotating crystal island.

Because the crystal orientation changes with position, the Kikuchi pattern observed in the CCD image changes slowly and continuously with the motion of the electron landing position on the rotating crystal island. This is because the resolution of the currently adopted EBSD equipment is about 10 nm, so the angular rotation of the crystal over 10 nm on the rotating crystal island is only 0.2°. At such a small resolution, the Kikuchi pattern generated at each local electron landing position can be considered the Kikuchi pattern measured at a single $In_2O_3$ crystal at a certain orientation. Thus for an ideal disk-shaped rotating island crystal, such a special Kikuchi pattern is observed when electrons land on its center.

Interestingly, this phenomenon is a physical manifestation of the Banach fixed point theorem in mathematics. The famous Banach fixed point theorem [89] is an important tool that guarantees the existence and uniqueness of fixed points of certain self-maps of metric spaces and provides a constructive method to find those fixed points. The Banach fixed point theorem is also known as the contraction mapping theorem or contractive mapping theorem.

An example is that if the map of a country is printed at a reduced size inside the territory of that country, there is one and only one point on the map whose position indicates the location of the land on which it falls. Here, the Kikuchi pattern observed in the rotating crystal sample SEM image must coincide with one of the Kikuchi patterns in the upper region of the CCD image taken at a certain location in this rotating crystal sample in the EBSD Kikuchi pattern matrix map. It is very interesting to note that this small Kikuchi pattern region in the observed CCD image, which originates from electrons emitted near the surface normal of the rotating crystal island, shows almost the same pattern as the Kikuchi patterns observed in the BSE and SE images (Fig. 5c and 5d). This is an example of the Banach fixed point theorem.

Notably, almost the same Kikuchi pattern can be observed in both the SE and BSE images, although the SE image's contrast is significantly lower than the BSE image. This result implies that the diffraction information observed in SE images does not come from the interaction when the SEs are transported inside the crystal. It instead comes from the cascade SEs produced by the backscattered electrons that are yield modulated by the incident high energy electrons interacting with the crystal lattice. Therefore, the quantities for SEs emitted from the different regions of the crystal surface are influenced by the local crystal direction, where primary electrons land on different regions of the rotational crystal.

It is worth mentioning that previous reports have observed many rotating crystals that share the property that the crystal orientation at the crystal's local position rotates depending on the position. However, most of these reported rotating crystals differ in both physical properties and appearance. Therefore, a more detailed classification of these special crystals is required based on the essential differences in how their local crystal orientation rotates. As in the case of crystalline materials, they are classified according to the different symmetries they exhibit. We advocate applying similar ideas to classify rotating crystals in a more refined way. For example, the two-dimensional rotating crystalline film found in this paper not only satisfies the basic characteristics of a rotating crystal (i.e., the crystal orientation at the local position of the crystal rotates depending on the position) but also satisfies cylindrical symmetry with respect to the crystal orientation. Therefore, the rotating crystals found in this paper is named cylindrically symmetric rotating crystals.

Finally, it is reiterated that the exceptional cylindrically symmetric rotating crystal thin film prepared in this paper possesses physically interesting properties. This film provides a very extreme case study of SEM contrast. In the SEM raster scan mode, the cylindrically symmetric rotating crystal film demonstrates the particle properties of the incident electrons in the image, i.e., the surface morphology of the sample detected by the interaction between incident electrons and the sample. However, it also exhibits the wave properties of the incident electrons, i.e., the lattice-related information of the sample carried by the diffraction interaction of the incident electrons with the sample lattice.

In addition, the special cylindrically symmetric rotating crystal film prepared in this paper is also mathematically interesting. The SEM image of this crystal film is mathematically associated with the EBSD Kikuchi pattern map, providing a physical example of the Banach fixed point theorem in materials science. For this crystal, the SEM image showing the Kikuchi pattern contrast can be found in the almost identical corresponding EBSD Kikuchi pattern map. In fact, this Banach fixed point-like relationship between the SEM image and corresponding EBSD Kikuchi pattern map for the cylindrically symmetric rotating crystal is guaranteed by the time-reversal symmetry, which is widely used in EBSD theoretical simulations. However, it has appeared here in experimental EBSD measurements in an interesting way.

## Methods

**Rotating crystal film preparation.** Amorphous films with a 30 nm thickness were prepared by DC magnetron sputtering (Shibaura Mechatronics, CFS-4EP-LL i-Miller) on a sapphire substrate at room temperature. Sputtering targets consisting of $In_2O_3$ and $SiO_2$ were used. The ratio of Si/In in the sputtering target was 2.3 at. %, corresponding to 1 wt. % in terms of $SiO_2/(In_2O_3 + SiO_2)$. The sputtering target and the substrate were separated by 160 mm in the sputtering system. InSiO films were generated by a plasma at 200 W under a mixed atmosphere of argon and oxygen at a 1:1 gas flow ratio and 0.25 Pa total pressure. This gas flow ratio was determined to produce electrically stable InSiO films resistant to thermal stress [49,90,91]. In situ SEM observations were performed to obtain images of dynamic crystallization along a precisely fixed observation area. The InSiO films were heated in situ to 300 °C in the SEM environment. Figure 2a shows typical results of SEM and EBSD observations of InSiO films after crystallographic heating.

**SEM measurements.** In situ SEM observations were performed by combining the Gatan Murano heating sub-stage and Hitachi SU5000 Schottky SEM. This combined system allowed us to obtain images of crystallization dynamically along a precisely fixed observation area.

**EBSD measurements.** Diffraction patterns were collected using a JSM-IT800 SEM equipped with a CCD detector in high resolution (1244×1024) mode. The setup geometry was held constant with a 10.0±0.1 mm working distance. The imaging parameters were a 15 kV accelerating voltage, ~20 nA beam current, and 0.1 ms dwell time. After collecting high-resolution EBSDs from each material, all patterns collected were exported as TIF images.

[38] Hirsch, P. B., Howie, A., Nicholson, R. B., Pashley, D. W., & Whelan, M. J. *Electron Microscopy of Thin Crystals* (Butterworths, London 1977).

[39] Kikuchi, S. Electron diffraction in single crystals. *Jpn. J. Appl. Phys.* **5**, 83-96 (1928).

[40] Coates, D. G. Kikuchi-like reflection patterns obtained with the scanning electron microscope. *Philos. Mag.* **16**, 1179-1184 (1967).

[41] Joy, D. C., Newbury, D. E. & Davidson, D. L. Electron channeling patterns in the scanning electron microscope. *J. Appl. Phys.* **53**, R81-R122 (1982).

[42] Venables, J. A. & Harland, C. J. Electron backscattering patterns—A new technique for obtaining crystallographic information in the scanning electron microscope. *Philos. Mag.* **27**, 1193-1200 (1973).

[43] Langer, E. & Däbritz, S. Investigation of HOLZ rings in EBSD patterns. *Phys. Status Solidi C* **4**, 1867-1872 (2007).

[44] Venables, J. A. & Spiller, G. D. T. Nucleation and growth of thin films. *Surface Mobilities on Solid Materials* 341-404 (Springer, Boston, MA, 1983).

[45] Evans, J. W., Thiel, P. A. & Bartelt, M. C. Morphological evolution during epitaxial thin film growth: Formation of 2D islands and 3D mounds. *Surf. Sci. Rep.* **61**, 1-128 (2006).

[46] Gonzalez, D., Kelleher, J. F., da Fonseca, J. Q. & Withers, P. J. Macro and intergranular stress responses of austenitic stainless steel to 90 strain path changes. *Mater. Sci. Eng. A*, **546**, 263-271 (2012).
32

**Acknowledgements** This work was supported by the National Institute for Materials Science under the Support system for curiosity-driven research, JSPS KAKENHI Grant Number JP21K14656, Grant for Basic Science Research Projects from The Sumitomo Foundation and from The Kao Foundation for Arts and Sciences. J.H. was supported by the U.S. Department of Energy (DOE), Office of Science, Office of Basic Energy Sciences under Award DE-SC0019467. Z.J.D was supported by the National Key Research and Development Project (2019YFF0216404) and Education Ministry through "111 Project 2.0" (BP0719016). All DFT calculations were performed on the Numerical Materials Simulator supercomputer at the National Institute for Materials Science. We are thankful for helpful discussions and suggestions from Dr. Takio Kizu.

**Author Contributions** B.D., K.T., Z.J.D. supervised the project. B.D. designed the research. B.D., X.L. wrote the manuscript with important input from all authors. B.D., K.S. performed the experiments. L.C. performed the calculations. All authors discussed the results and commented on the manuscript.

**Competing financial interests** The authors declare no competing financial interests.

## Supporting materials

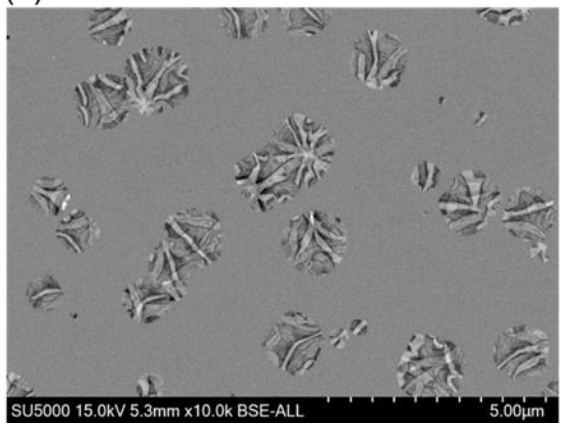 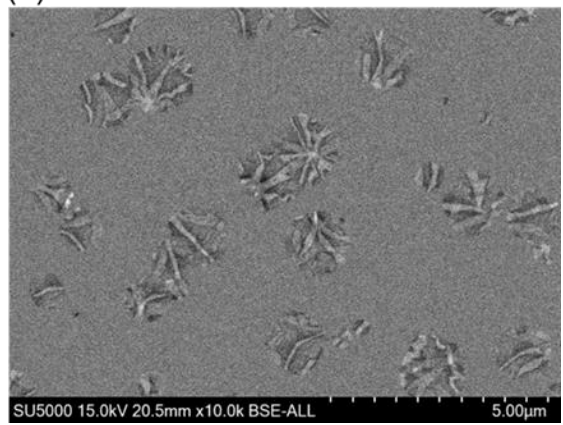

**Fig. S1 SEM images at different working distances.** BSE images were taken with a working distance of: **a** 5.3 mm; **b** 20.5 mm. The figures are for the BSE of an InSiO film, where the rotational crystal islands are observed after annealing at 300 °C.

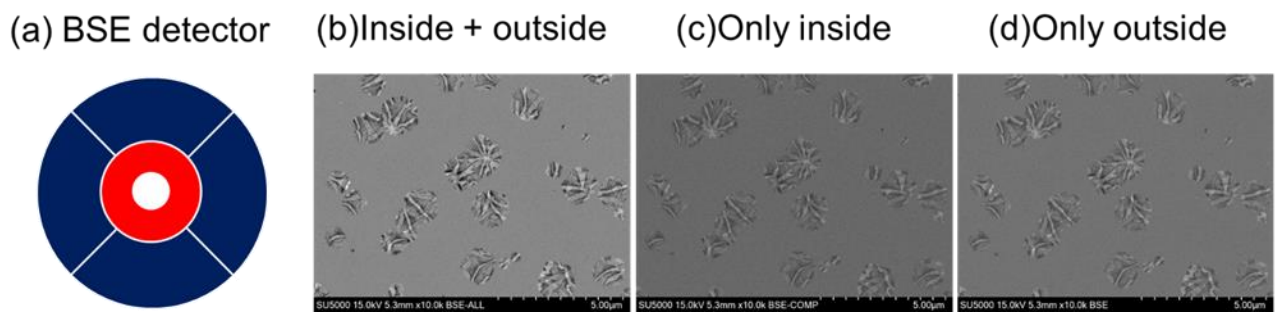

**Fig. S2 a** Structure of the BSE detector. **b–d** BSE images taken using the different parts of the BSE detector. The red part is the detector inside, while the blue part is the outside. The rotational crystal islands are observed after annealing at 300 °C.

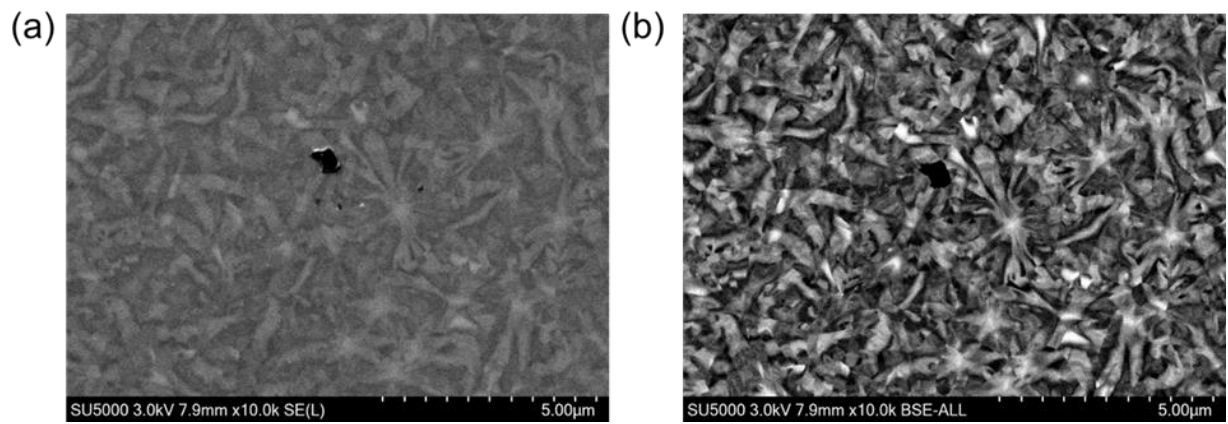

**Fig. S3 a** SE, and **b** BSE, images for the final annealing state at a temperature of 300 °C and electron energy of 3 keV.

**Mov. S1** The mp4 file shows the crystallization process of InSiO film from 250 °C to 300 °C at an incident electron energy of 15 keV.